\documentclass{aa}  

\usepackage{graphicx}
\usepackage{xcolor}
\usepackage{txfonts}

\usepackage{amssymb}
\usepackage{amsmath}
\usepackage{pict2e}
\usepackage{multirow}
\usepackage{dsfont}

\def\bf{\mbox{\boldmath$F$}}
\def\bu{\mbox{\boldmath$u$}}
\def\bxi{\mbox{\boldmath$\xi$}}
\def\br{\mbox{\boldmath$r$}}
\def\bx{\mbox{\boldmath$x$}}
\def\bs{\mbox{\boldmath$s$}}
\def\sc{\mbox{\boldmath$r$}_{\rm \! s}}
\def\zs{z_{\rm  s}}
\def\rs{r_{\rm  s}}
\def\ii{{\rm i}}
\def\EE{\mathds{E}}
\def\Var{{\rm Var}}
\def\dd{{\rm d}}

\begin{document}

\title{Signal and noise in helioseismic holography}

\author{
	Laurent Gizon\inst{1,2,3}
        \and Damien Fournier\inst{1}
         \and Dan Yang\inst{1}
          \and Aaron C. Birch\inst{1}
          \and H\'el\`ene Barucq\inst{4}
          }
          
\institute{
	Max-Planck-Institut f\"ur Sonnensystemforschung, Justus-von-Liebig-Weg 3, 37077 G{\"o}ttingen,
      	Germany \\ \email{gizon@mps.mpg.de}
      	\and
        Institut f\"ur Astrophysik, Georg-August-Universit{\"a}t G\"ottingen, Friedrich-Hund-Platz 1, 37077 G{\"o}ttingen, Germany
        \and
        Center for Space Science, NYUAD Institute, New York University Abu Dhabi, PO Box 129188, Abu Dhabi, UAE
        \and  
        Magique-3D, Inria Bordeaux Sud-Ouest, E2S UPPA, 64000 Pau, France 
        }

   \date{Received XXX; accepted  XXX}

% \abstract{}{}{}{}{} 
% 5 {} token are mandatory
 
  \abstract{
  % context heading (optional)
  % leave it empty if necessary  
   Helioseismic holography is an imaging technique used to study heterogeneities and flows in the solar interior from observations of solar oscillations at the surface. Holograms contain noise due to the stochastic nature of solar oscillations.}{
  % aims heading (mandatory)
   We provide a theoretical framework for modeling signal and noise in Porter-Bojarski helioseismic holography. }{
  % methods heading (mandatory)
   The wave equation may be recast into a Helmholtz-like equation, so as to connect with the acoustics literature and define the holography Green's function in a meaningful way. Sources of wave excitation are assumed to be stationary, horizontally homogeneous, and  spatially uncorrelated. Using the first Born approximation we calculate holograms in the presence of perturbations in sound-speed, density, flows, and source covariance, as well as the noise level as a function of position. This work is a direct extension of the methods used in time-distance helioseismology to model signal and noise.  }{
  % results heading (mandatory)
   To illustrate the theory, we compute  the hologram intensity numerically for a buried sound-speed perturbation at different depths in the solar interior. The reference Green's function is obtained for a spherically-symmetric solar model using a finite-element solver in the frequency domain.
   Below the pupil area on the surface, we find that the spatial resolution of the hologram intensity is very close to half the local wavelength. For a sound-speed perturbation of size comparable to the local spatial resolution,  the signal-to-noise ratio is approximately constant with depth.  Averaging the hologram intensity over a number $N$ of frequencies above 3~mHz  increases the signal-to-noise ratio by a factor nearly equal to the square root of $N$. This may not be the case at lower frequencies, where large variations in the holographic signal are due to the individual contributions of the long-lived  modes of oscillation. }{}


   \maketitle
%
%________________________________________________________________

\section{Introduction}

Local helioseismology aims at studying the solar interior in three dimensions by exploiting the information contained in the waves observed at  the solar surface \citep[e.g.,][and references therein]{GIZ05}. Helioseismic holography is one particular approach of local helioseismology, which images subsurface scatterers by back-propagating the surface wave field to target points in the interior. Helioseismic holography is also known as Lindsey-Braun (LB) holography \citep[][and references therein]{LIN97, LIN00a}. It has been used to study solar convection \citep{Braun2004,Braun2007},  active region emergence \citep{Birch2013,Birch2016}, sunspot subsurface structure  \citep{Braun2008,Birch2009}, to image wave sources \citep{Lindsey2006}, to study sunquakes caused by solar flares \citep{ZHA13,BES17}, and to detect active regions on the far side of the Sun  \citep{LIN00b,Liewer2014}.

In acoustics, a well-established version of holography in a medium that contains  sources is Porter-Bojarski (PB) holography \citep{POR82}. PB holography uses both the wave field and its normal derivative at the surface to produce holograms \citep{POR69}. PB holography was introduced in helioseismology by \citet{SKA01,SKA02}, where deterministic sources and scatterers were recovered in a solar background. \citet{YAN18} recently studied PB holograms in a homogeneous medium permeated by localized deterministic sources to study ghost images near the observational boundary.

In this paper we apply PB holography in a realistic helioseismological setting. First we rewrite the wave equation in Helmholtz form, in order to properly define the Green's functions that are involved in the definition of PB holograms. The background density and sound-speed are taken from a standard solar model. The model of wave excitation is described by a reasonable source covariance function, which leads to a solar-like power spectrum for acoustic oscillations. 

The signal is defined as the expectation value of the perturbations in hologram intensity that result from perturbations in sound speed, density, and flows with respect to the reference solar model. The corresponding sensitivity kernels are computed in the first-order Born approximation \citep{GIZ02, Birch2007, Braun2007, Birch2011}. This signal must take into account the correlations between incident and scattered wave fields, which are both connected to the sources of excitation (turbulent convection).

Random noise in holograms is due to the stochastic nature of the sources of  excitation. While noise can sometimes be estimated from the data \citep{LIN90,BRA08}, a theoretical understanding is useful to design holographic experiments. Here we extend to holography the noise model developed in time-distance helioseismology \citep{GIZ04,FOU14}. We do not attempt to image individual sources as in \citet{SKA02}, which in our view is not a well-posed problem \citep[see also][]{Lindsey2006}{, except in case of imaging the sources of sunquake waves}. Instead we consider sources to be specified through a source covariance function.

\section{Reduced wave equation}
At angular frequency $\omega$ and spatial position $\br$ in the computational domain $V$, the propagation of acoustic waves in a 3D heterogeneous moving medium is described by the displacement vector $\bxi(\br,\omega)$, solution to 
\begin{equation}
- (\omega + \ii \gamma + \ii \bu \cdot\nabla)^2\bxi  -  \frac{1}{\rho}  \nabla\left( \rho c^2 \nabla\cdot\bxi \right) + \text{gravity  terms}  = \bf,
\end{equation}
where  $\rho(\br)$ and $c(\br)$ are the density and sound speed, and $\bu(\br)$ is a steady vector flow. Wave attenuation is included through the function $\gamma(\br,\omega)$. The source term $\bf(\br,\omega)$ is a realization from a random process; it describes the stochastic excitation of the waves by turbulent convection. Following \citet{LAM09} and \citet{DEU84}, we consider the scalar variable 
\begin{equation}
\psi =  \rho^{1/2} c^2 \nabla\cdot\bxi,
\end{equation} 
to recast the wave equation into a Helmholtz-like equation
\begin{equation}
L\psi := - (\Delta + k^2) \psi -  \frac{2\ii \omega}
{\rho^{1/2}c} \rho\bu \cdot \nabla \left( \frac{ \psi } { \rho^{1/2}c} \right)
= S, \label{eq:wave}
\end{equation}
where  $S = \rho^{1/2} c^2 \nabla\cdot\bf$ is a scalar source term. The local wavenumber $k(\br,\omega)$ is given by
\begin{equation}
k^2 = \frac{(\omega^2 + 2\ii \omega \gamma) - \omega_{\rm c}^2}{c^2}, \label{eq:k}
\end{equation}
where the squared acoustic cut-off frequency is
\begin{equation}
\omega_{\rm c}^2 = \rho^{1/2} c^2 \Delta ( \rho^{-1/2} ).
\end{equation}
{In obtaining Eq.~(\ref{eq:wave}),} we ignored gravity terms and assumed slow variations of $c$, $\bu$, and $\gamma$ compared to the wavelength \citep{GIZ17}. The advection term is such that the corresponding operator is Hermitian symmetric for the inner product 
$\langle \psi_1, \psi_2 \rangle = \int \psi_1^*  \psi_2  \, dV$ under the  conditions that the flow conserves mass and that it does not cross the boundary ($u_n =0$ on $\partial V$).

The stochastic sources of excitation are assumed to be stationary and spatially uncorrelated. They are described by a source covariance function of the form
\begin{equation}
\EE [ S^*(\br, \omega) S(\br',\omega) ] = M(\br, \omega) \delta(\br-\br').
\end{equation}

To solve Eq.~\eqref{eq:wave}, one needs to specify a boundary condition at the computational boundary. Here we apply an outgoing radiation boundary condition 
\begin{equation}
\partial_n \psi = \ii k_n \psi \quad \textrm{ on } \partial V . \label{eq:b.c.}
\end{equation}
We apply the boundary condition (Atmo RBC 1) from  \citet{BAR17}, which assumes an exponential decay of the {background} density at the boundary of the domain but neglects curvature. Then, the local wavenumber $k_n$ from Eq.~\eqref{eq:b.c.} is given by
\begin{equation}
k_n^2 = \frac{\omega^2+2\ii\omega\gamma}{c^2}  -  \frac{1}{4H^2}, \label{eq:kn}
\end{equation}
where $H = -1/( {\textrm{d} \ln \rho} / {\textrm{d} r}  )$ is the density scale height at the boundary.
The last term in Eq.~\eqref{eq:kn} is connected to the cut-off frequency for an isothermal atmosphere \citep{LAM09}, thus $k_n$ is an approximation of the wavenumber $k$ from  Eq.~\eqref{eq:k}. \citet{FOU17} discusses several of the boundary conditions derived in \citet{BAR17}.

%%%%%%%%%%%%%%%%%%%
\section{Hologram and hologram intensity}
%%%%%%%%%%%%%%%%%%%

The following calculations are done at constant $\omega$, thus we drop $\omega$ from the list of function arguments  when not explicitly needed.
The Porter-Bojarski hologram is defined by \citet{POR82}:
\begin{equation}
\Phi_\alpha (\bx, A) := \int_{A}  [\psi (\br) \partial_n  H_\alpha (\br, \bx) - H_\alpha (\br, \bx)  \partial_n \psi(\br)]  \dd \br , \label{eq:hologram}
\end{equation}
where $H_\alpha$ is an acoustic wave propagator for the reference medium and {$A$ is a surface on the Sun where  $\psi$ and $\partial_n \psi$ are observed}. The role of $H_\alpha$ is to propagate the wave field backward (or forward) in time, which leads to the concept of egression (or  ingression) in LB  holography \citep{LIN00a}. 

\begin{table}[t]
\caption{Possible wave propagators.}
\begin{tabular}{ccc}
\hline \hline
Wave propagator & $H_\alpha$ &  References \\ \hline
 {Backward} & $G_0^\ast$  & \cite{POR82} \\
{Backward} & $G^-_0$  & \cite{TSA87} \\ 
{Backward}  & Im $G_0$ &   \cite{DEV85} \\ 
Forward  & $(G^-_0)^\ast$ & This work \\ \hline
\end{tabular}
\label{tab:wavePropagator}
\end{table}

Several choices have been proposed in the literature for the propagators, as detailed in Table~\ref{tab:wavePropagator}. 
These depend on the outgoing ($G_0$) and incoming ($G_0^-$) Green's functions defined in a reference medium with $\rho_0$, $c_0$, $\gamma_0$, and $\bu_0= 0$:
\begin{eqnarray}
L_0 [ G_0 (\br, \br') ] = \delta(\br - \br') & \text{and} & \partial_n G_0 = \ii k_n G_0 \textrm{ on } \partial V,
\label{Greens}
\\
L_0 [ G^-_0 (\br, \br') ] = \delta(\br - \br')  &\text{and}& \partial_n G^-_0 
= - \ii k_n G^-_0   \textrm{ on } \partial V,
\end{eqnarray}
with 
\begin{equation}
L_0 = - (\Delta + k_0^2) ,
\end{equation}
where $k_0$ is $k$ in the reference medium and $k_n$ is from Eq.~\eqref{eq:kn}.
The Green's functions $G_0^\ast$ or $G_0^-$ are backward propagators (c.f. egression), while  $(G_0^-)^\ast$ is a forward  propagator (c.f.  ingression). When the surface $A$ is closed, it is equivalent to use $G_0^\ast$ and $\textrm{Im} \ G_0$ \citep{DEV85}.
\citet{TSA87} proposed $H_\alpha  = G_0^-$ as a backward propagator to correct  for wave attenuation. 

If the observations are made at the computational boundary and the wave field satisfies the same boundary condition as the Green's function, then Eq.~\eqref{eq:hologram} reads
\begin{equation}
\Phi_\alpha (\bx, A) = \int_{A}  \psi (\br) [ \partial_n  H_\alpha (\br, \bx) - \ii k_n H_\alpha (\br, \bx)]  \dd \br.
\end{equation}
When $H_\alpha = G_0^\ast$, we have
\begin{equation}
\Phi (\bx, A) = -2\ii \textrm{Re}[k_n] \int_{A}  \psi (\br) G_0^\ast(\br, \bx)  \dd \br,
\end{equation}
which corresponds to the egression as defined by \citet{LIN00a}. Thus LB and PB holograms are closely related{, at least for the upper boundary condition employed here}.

\begin{figure}[t]
\begin{center}
\includegraphics[width=0.8\linewidth]{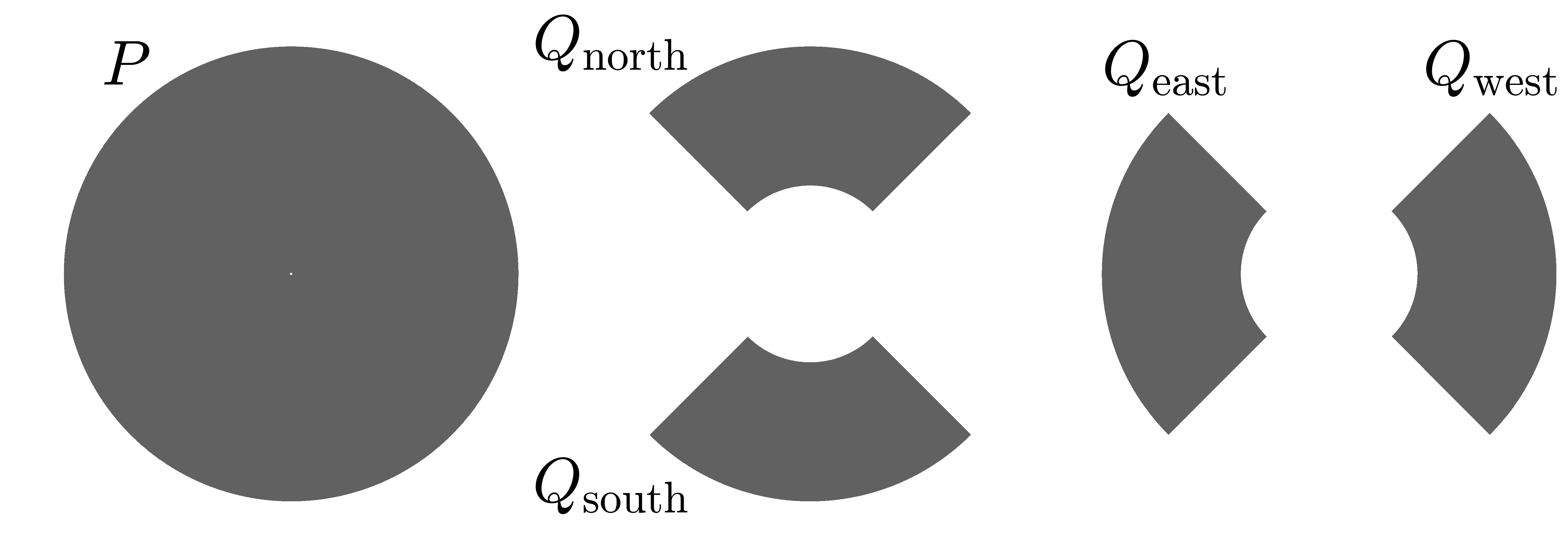} 
\end{center}
\caption{Pupil geometries used to compute sound-speed or source kernels ($P$) and flow kernels ($Q$s), also see  Table~\ref{tab:pupils}.} \label{fig.1}
\end{figure}

In LB holography, information is extracted from the egression-ingression correlation (wave-speed perturbations and flows) and from the egression power (source covariance).  Analogously, we define the PB hologram intensity (or hologram covariance) as
\begin{equation}
I_{\alpha\beta}(\bx, A, A') =  \Phi^*_\alpha(\bx, A)  \Phi_\beta (\bx, A')  .
\end{equation}
For the case $\alpha=\beta$ we define
\begin{equation}
I_{\alpha}(\bx, A) =  |\Phi^*_\alpha(\bx, A)  |^2  .
\end{equation}
Different choices of pupils and propagators will provide sensitivity to different quantities as shown in Table~\ref{tab:pupils}. Scatterers are detected by correlating forward and backward propagated holograms. Different pupil shapes will give access to different components of the flow (Table~\ref{tab:pupils} and Fig.~\ref{fig.1}).

\begin{table}[t]
\caption{Proposed propagators and associated pupils $(H_\alpha, A)$  and $(H_\beta, A')$ for different types of perturbations.
The pupil geometries are shown in Fig.~\ref{fig.1}.}
\begin{center}
\begin{tabular}{c cc cc}
\hline \hline  
Perturbations & $H_\alpha$ & $A$ & $H_\beta$ &  $A'$  \\
\hline 
Source covariance         &$\textrm{Im} \ G_0$ & $P$ &  $\textrm{Im} \ G_0$   &  $P$  \\
Sound speed     & $(G^-_0)^\ast$  & $P$ & $G_0^-$ & $P$    \\
Flow $u_\theta$ & $(G^-_0)^\ast$ & $Q_{\rm south}$& $G_0^-$ & $Q_{\rm north}$   \\
Flow $u_\phi$   &$(G^-_0)^\ast$  & $Q_{\rm east}$ & $G_0^-$ & $Q_{\rm west}$  \\
 \hline
\end{tabular}
\label{tab:pupils}
\end{center}
\end{table}

\section{First-order perturbations with respect to a reference solar model}

We wish to study how perturbations to the background medium {affect} holograms. Using the Born approximation, the first step is to express the perturbations to the wavefield and use this expression in Eq.~\eqref{eq:hologram} to obtain the perturbations to the hologram and hologram intensity.

\subsection{Perturbations to the wavefield}

 We consider perturbations $\delta c$, $\delta \rho$, $\delta \gamma$,  $\bu$ with respect to the reference medium defined by $\rho_0$, $c_0$, $\gamma_0$,  and $\bu_0=0$. The perturbations to the sources of excitation  are described through the perturbations to the source covariance,
\begin{equation}
M (\br,\omega)= M_0  (\br,\omega)+ \delta M(\br,\omega),
\end{equation}
where
\begin{equation}
\delta M  (\br,\omega)= \EE[S_0^*(\br) \delta S(\br')+ \delta S^*(\br) S_0(\br')] =  \epsilon(\br) M_0 (\br,\omega).
\end{equation}
Using the Born approximation up to first order, we write the wave field as
\begin{equation}
\psi (\br, \omega) = \psi_0(\br, \omega) + \delta \psi(\br, \omega) ,
\end{equation}
where the zeroth- and first-order wave fields are given by
\begin{align}
L_0 [\psi_0] & = S_0    , 
\label{eq.zero} \\
L_0 [\delta \psi]  &= - \delta L [\psi_0] + \delta S 
. \label{eq.first}
\end{align}
The perturbed wave operator is 
\begin{equation}
 \delta L  [\psi_0] = - \delta k^2 \psi_0 - \frac{2\ii \omega}
{\rho^{1/2}_0c_0} \rho_0\bu \cdot \nabla \left( \frac{ \psi_0 } { \rho^{1/2}_0c_0} \right),
\end{equation}
with
\begin{equation}
\label{ksquared}
c_0^2\, \delta k^2 = {2\ii\omega\; \delta \gamma}  - (\omega^2+2\ii \omega\gamma_0) \frac{\delta c^2}{c_0^2} -  \left(\frac{\partial \omega_{\rm c}^2}{\partial\rho}  \right)_0 \delta \rho.
\end{equation}

In terms of the Green's function $G_0$, we have
\begin{align}
  & \psi (\br) = \psi_0(\br) +  \int_V   G_0(\br, \sc)   \delta k^2(\sc) \psi_0(\sc)  \dd  \sc  + \int_V   G_0(\br, \bs)   \delta S (\bs)  \dd \bs \nonumber \\
  &+ 2\ii \omega \int_V   \rho_0(\sc)\, \bu_0(\sc) \cdot \nabla \left( \frac{G_0(\br, \sc)}{\rho_0^{1/2}(\sc) c_0(\sc)} \right) \frac{\psi_0(\sc)}{\rho_0^{1/2}(\sc) c_0(\sc)}  \dd \sc. \label{eq:dpsi}
\end{align}
where the scattering location $\sc$   spans  the entire volume $V$. 

\subsection{Perturbations to the hologram}
For convenience, we introduce the source kernels
\begin{equation}
K_\alpha (\bx,\bs, A) =  \int_{A}  [G (\br,\bs) \partial_n  H_\alpha (\br, \bx) - H_\alpha (\br, \bx)  \partial_n G(\br,\bs)]  \dd \br \; ,
\end{equation}
such that the PB hologram is given by
\begin{equation}
\Phi_\alpha (\bx, A) = \int_V  K_\alpha (\bx,\bs, A) S(\bs) \dd \bs. \label{eq:wavefield}
\end{equation}
We denote by $K_{\alpha,0}$ the source kernel in the reference medium (when $G$ is replaced by $G_0$).

Replacing the wavefield  by its first order expansion (Eq.~\eqref{eq:dpsi}) in the definition of the hologram (Eq.~\eqref{eq:hologram}), one obtains the perturbation to the hologram
\begin{equation}
\Phi_\alpha (\bx, A) = \Phi_{\alpha,0} (\bx,A) + \delta\Phi_\alpha (\bx,A),
\end{equation}
where $\Phi_{\alpha,0}$ is the hologram in the background medium given by Eq.~\eqref{eq:wavefield} when $K_\alpha$ is replaced by $K_{\alpha,0}$ and $\delta \Phi_\alpha$ expresses the changes in the hologram due to the perturbations from the background medium 
\begin{equation}
\delta \Phi_{\alpha} (\bx,A) = \int_V \delta K_\alpha (\bx,\bs, A) S(\bs) \dd \bs  + \int_V  K_{\alpha,0} (\bx,\bs, A) \delta S(\bs) \dd \bs  ,
\end{equation}
where 
\begin{align}
\delta K_\alpha &(\bx,\bs, A) =   \int_V K_{\alpha,0}(\bx,\sc, A) \delta k^2(\sc)  G_0(\sc ,\bs)    
\dd \sc  \nonumber \\
 &+   2\ii \omega \int_V \frac{K_{\alpha,0}(\bx,\sc, A)}{\rho_0^{1/2}(\sc) c_0(\sc)}  \rho_0 \bu_0 \cdot \nabla \left(  \frac{G_0(\sc ,\bs)}{\rho_0^{1/2}(\sc) c_0(\sc)} \right)    \dd \sc.
\end{align}

\subsection{Perturbations to  hologram intensity}

To first order, we write the hologram intensity in the form
\begin{equation}
 I_{\alpha \beta}(\bx, A, A') = I_{\alpha \beta, 0}(\bx, A, A') + \delta I_{\alpha \beta} (\bx, A, A') .
 \end{equation}
The expectation values of the zeroth- and first-order hologram intensities are
 \begin{align}
 \EE  [I_{\alpha \beta, 0}(\bx, A, A')] &=
 \int_V K^*_{\alpha,0}(\bx,\bs, A) K_{\beta,0}(\bx,\bs, A')  M_0(\bs)  \dd \bs   ,
 \\  
  \EE[ \delta I_{\alpha\beta}(\bx, A, A') ]&=   \int_V  K_{\alpha,0}^*(\bx,\bs, A)  \delta K_{\beta}(\bx,\bs, A')  M_0(\bs)  \dd \bs 
 \nonumber \\ 
  & \hspace*{0.5cm} +     \int_V   \delta K_{\alpha}^*(\bx,\bs, A) K_{\beta,0}(\bx,\bs, A')    M_0(\bs)  \dd \bs \nonumber 
\\
&\hspace*{0.5cm} +  \int_V K_{\alpha,0}^*(\bx,\bs, A)  K_{\beta,0}(\bx,\bs, A') \epsilon(\bs) M_0(\bs)  \dd \bs.   
\end{align}
Using the definition of the source kernels, we obtain 
\begin{equation}
\EE [I_{\alpha\beta,0} (\bx, A, A')] 
= \langle \!\langle   C_0(\br,\br')  \rangle\!\rangle_{\alpha \beta}(\bx, A, A') \label{eq:I0} ,
 \end{equation}
 where 
 \begin{equation}
C_0(\br,\br') = \int_V G_0^*(\br,\bs) G_0(\br',\bs) M_0(\bs) \dd \bs \label{eq:C0}
\end{equation}
is the expectation value of the cross-covariance function and, for any function $F(\br,\br')$, 
the double brackets mean
 \begin{align}
 \langle \!\langle   F(\br,\br')  \rangle\!\rangle_{\alpha \beta}(\bx, A,A') & = 
 \nonumber \\ 
\int_A \dd \br \int_{A'} \dd \br' \;  \biggl[&
\partial_n  H^*_\alpha (\br,\bx) F(\br,\br')  \partial_{n'} H_\beta (\br',\bx) \nonumber 
\\
&
-    H^*_\alpha (\br,\bx)
\partial_n F(\br,\br')  
\partial_{n'}H_\beta (\br',\bx) \nonumber
\\
&
-    \partial_{n}H^*_\alpha (\br,\bx) 
\partial_n F(\br,\br')  
H_\beta(\br',\bx)  \nonumber
\\
&  
 + 
H^*_\alpha (\br,\bx)  \, \partial_{n} \partial_{n'}F(\br,\br') \, H_\beta (\br',\bx)   \biggr] .  
\end{align}
The perturbation to the hologram intensity is given by
\begin{align}
& \EE [\delta I_{\alpha\beta} (\bx, A, A')] =  \int_V \epsilon(\bs) \mathcal{K}_{\alpha \beta}^\epsilon( \bx,\bs, A, A') \, \dd \bs \nonumber \\
& +
\int_V    \left( \delta k^{2*}(\sc) \,  \mathcal{K}^k_{\alpha\beta}(\bx,\sc, A, A') + \delta k^{2}(\sc) \,  \mathcal{K}^{k*}_{\beta\alpha}(\bx,\sc, A', A) \right) \, \dd \sc
\nonumber\\
& + 2\ii \omega \int_V  \bu(\sc)  \cdot  \left( \boldsymbol{ \mathcal{K}}^u_{\alpha\beta}(\bx,\sc, A, A') - \boldsymbol{\mathcal{K}}^{u \ast}_{\beta\alpha}(\bx,\sc, A', A) \right) \, \dd \sc, 
\end{align}
where
\begin{align}
\mathcal{K}_{\alpha \beta}^\epsilon&(\bx,\bs, A, A') =M_0(\bs) 
\left\langle\!\left\langle G_0^\ast(\br,\bs) G_0(\br',\bs)  \right\rangle\!\right\rangle_{\alpha \beta}(\bx,A,A'), \\
 \mathcal{K}^k_{\alpha\beta}&(\bx,\sc, A, A')  = 
\left\langle\!\left\langle C_0(\sc,\br')
 G_0^\ast(\br, \sc)     \right\rangle\!\right\rangle_{\alpha \beta}(\bx,A,A'), \label{eq:Kk2} \\
 \boldsymbol{\mathcal{K}}_{\alpha\beta}^{u}&(\bx, \sc,A, A')  = \nonumber \\
& \left\langle\!\left\langle \nabla \left( \frac{C_0(\sc,\br')}{\rho_0(\sc)^{1/2} c_0(\sc)} \right)
 \frac{\rho_0(\sc)^{1/2} G_0^\ast(\br, \sc) }{c_0(\sc)}    \right\rangle\!\right\rangle_{\alpha \beta}(\bx,A,A').
 \end{align}
 
The kernels for $\delta k^2$ and $\delta k^{2*}$ can be combined using Eq.~(\ref{ksquared}) to obtain kernels for sound-speed $\mathcal{K}_{\alpha\beta}^c$, density $\mathcal{K}_{\alpha\beta}^\rho$ and attenuation $\mathcal{K}_{\alpha\beta}^\gamma$.
For example, 
\begin{equation}
\EE [\delta I_{\alpha\beta} (\bx, A, A')] =  \int_V \delta c (\sc) \, \mathcal{K}_{\alpha \beta}^c(\bx,\sc, A, A') \, \dd \sc \label{eq:signalC}
\end{equation}
with
\begin{align}
 \mathcal{K}^c_{\alpha\beta} (\bx,\sc,A, A')=& 
 - \frac{2 (\omega^2-2\ii \omega \gamma)}{c_0^3(\sc)}  \mathcal{K}^{k}_{\alpha\beta}(\bx,\sc,A, A')   \nonumber \\ 
 & -\frac{2 (\omega^2+2\ii \omega \gamma)}{c_0^3(\sc)} \mathcal{K}^{k*}_{\beta\alpha}(\bx,\sc,A', A) 
.
\label{eq.Kc}
\end{align}

\subsection{Choice of the source covariance}

In order to be able to compute the above kernels, one still need to choose the source covariance function $M_0$ in order to define the reference cross-covariance $C_0$ using Eq.~\eqref{eq:C0}. One possibility is to place the sources at a single depth, a few hundred kilometers below the solar surface. 

Another possibility is to choose a source covariance of the form 
\begin{equation}
M_0(\br,\omega)  = \Pi(\omega) \frac{\gamma(\br, \omega)}{c_0^2(\br)}, 
\label{eq.convenient}
\end{equation}
where $\Pi(\omega)$ is the source power spectrum \citep[see][]{GIZ17}. This choice implies
\begin{equation}
\label{eq.Csurf}
C_0(\br,\br',\omega) 
= \frac{\Pi(\omega)}{2\omega} {\rm Im}\, G_0 (\br', \br, \omega)   + \text{ surface term}.
\end{equation}
The surface term depends on the boundary condition. It vanishes for a Dirichlet boundary condition (free surface), while it remains for radiative boundary conditions (e.g. Sommerfeld). 
Below the acoustic cutoff frequency, modes are trapped well below the observational and computational boundaries and the surface term is negligible.
In this paper we use Eq.~(\ref{eq.convenient}) in the convection zone and switch off the sources above the photosphere. By doing so, the surface term in Eq.~(\ref{eq.Csurf}) vanishes.

\section{Noise in holograms}

To compute the noise level, we compute the variance of the hologram intensity in the absence of scatterers:
\begin{align}
\label{eq.var}
\Var [ & I_{\alpha\beta,0}(\bx) ] =  
 {\rm Var}\left[  \int K^*_{\alpha,0}(\bx, \bs) K_{\beta,0}(\bx, \bs') S^*(\bs) S(\bs') \dd \bs \dd\bs' \right]
\nonumber \\
&=
 \int_{V^4}  K^*_{\alpha,0}(\bx, \bs_1)    K_{\beta,0}(\bx, \bs_1')  K_{\alpha,0}(\bx, \bs_2)    K^*_{\beta,0}(\bx, \bs_2')  \nonumber \\
&   \qquad \quad \times M_4(\bs_1, \bs_1', \bs_2, \bs_2') \, \dd\bs_1 \dd\bs_1'  \dd\bs_2 \dd\bs_2' 
\nonumber \\
&\quad -  \left| 
 \int_{V^2} K^*_{\alpha,0}(\bx, \bs) K_{\beta,0}(\bx, \bs') \EE \left[S^*(\bs)   S(\bs') \right] \dd\bs \dd\bs' \right|^2, 
\end{align}
where
\begin{equation}
M_4(\bs_1, \bs_1', \bs_2, \bs_2')
= \EE\left[ S^*(\bs_1)S(\bs_1')S(\bs_2)S^*(\bs_2') \right]. 
\end{equation}
Under the (very reasonable) assumption that $S$ is a realization drawn from  a Gaussian random process, the fourth-order moment is the sum of products of the second-order moments:
\begin{eqnarray}
M_4(\bs_1, \bs_1', \bs_2, \bs_2')
&=&
\EE \left[S^*(\bs_1)   S(\bs_1') \right]
\EE \left[S(\bs_2)   S^*(\bs_2') \right]
\nonumber  \\&+&
\EE \left[S^*(\bs_1)   S(\bs_2) \right]
\EE\left[S(\bs_1')   S^*(\bs_2') \right]
\nonumber  \\&+&
\EE \left[S^*(\bs_1)   S^*(\bs_2') \right]
\EE\left[S(\bs_2)   S(\bs_1')\right].
\end{eqnarray}
The first term in $M_4$  cancels out the squared term in Eq.~(\ref{eq.var}).
The third term is zero as the frequencies are uncorrelated.  Thus,
\begin{align}
\Var[ I_{\alpha\beta,0}(\bx) ] &=  \int_{V^4}  K^*_{\alpha,0}(\bx, \bs_1)    K_{\beta,0}(\bx, \bs_1')  K_{\alpha,0}(\bx, \bs_2)    K^*_{\beta,0}(\bx, \bs_2')  \nonumber \\
&  \quad \times \EE \left[S^*(\bs_1)   S(\bs_2) \right]
\EE \left[S(\bs_1')   S^*(\bs_2') \right] \dd\bs_1  \dd\bs_1'  \dd\bs_2 \dd\bs_2'  \nonumber 
\\
&= \int_V    
|K_{\alpha,0}(\bx, {\bs}) |^2    
  M({\bs}) \dd{\bs}  \int_V    
|K_{\beta,0}(\bx, {\bs}) |^2    
  M({\bs}) \dd{\bs}  \nonumber \\
  &=   \EE[ I_{\alpha,0}(\bx) ]  \EE[ I_{\beta,0}(\bx) ]  . \label{eq:noise}
\end{align}
When $\alpha=\beta$, the standard deviation of $I_{\alpha,0}$ is equal to its expectation value. 
This is because  the probability density function of $I_{\alpha, 0}$ is a $\chi^2$ with two degrees of freedom, i.e. an exponential distribution.

%%%%%%%%%%%%%%%%%%%%
\section{Average over frequencies and signal-to-noise ratio}
%%%%%%%%%%%%%%%%%%%%

In order to increase the signal-to-noise ratio, one usually averages the hologram intensity  over a range of frequencies $[\omega_0-\Delta\omega/2, \omega_0+\Delta\omega/2]$. For an observation duration $T$, this interval will contain $N= \Delta\omega \, T/2\pi$ independent frequencies.

The {frequency-averaged} perturbation to the hologram intensity (i.e.\ the signal) is  denoted by
\begin{equation}
\langle \delta I_{\alpha\beta}(\bx)\rangle
=  \frac{1}{N} \sum_{i=1}^N  \delta I_{\alpha\beta}(\bx, \omega_i).
\end{equation}

The variance of the noise in the average hologram intensity is then given by
\begin{align}
{\rm Var}  \left\langle I_{\alpha\beta, 0}(\bx)  \right\rangle &= {\rm Var} \left( \frac{1}{N} \sum_{i=1}^N   I_{\alpha\beta,0} (\bx,\omega_i)  \right) \\
& = \frac{1}{N^2}    \sum_{i=1}^N  {\rm Var}\ I_{\alpha\beta,0} (\bx,\omega_i)  \\
& = \frac{1}{N} \left\langle   {\rm Var}  \;  I_{\alpha\beta,0}  (\bx)  \right\rangle  ,
\end{align}
since the noise in holograms at different frequencies is uncorrelated.

The expected signal-to-noise ratio is thus
\begin{equation}
{\rm SNR}(\bx)  = 
\frac{ \left| \EE \langle \delta I_{\alpha\beta}(\bx) \rangle    \right|}{\sqrt{ {\rm Var}  \langle I_{\alpha\beta, 0}  (\bx)  \rangle}
}   
=  \frac{ \sqrt{N} \,  \left| \EE \langle \delta I_{\alpha\beta}(\bx) \rangle    \right|}{\sqrt{ \langle \EE [I_{\alpha,0}(\bx)] \,\EE [I_{\beta,0}(\bx)]\rangle  }  } .
\end{equation}
The number of available frequencies $N$ within a fixed frequency band $\Delta \omega$ is proportional to the observation duration $T$, hence the noise level goes like $T^{-1/2}$. Provided that  the frequency interval $\Delta\omega$ is small with respect to the variations of the signal, then the signal-to-noise ratio will increase like $T^{1/2}$.

\section{Example computations}

\begin{figure}[t]
\begin{center}
\includegraphics[width=\linewidth]{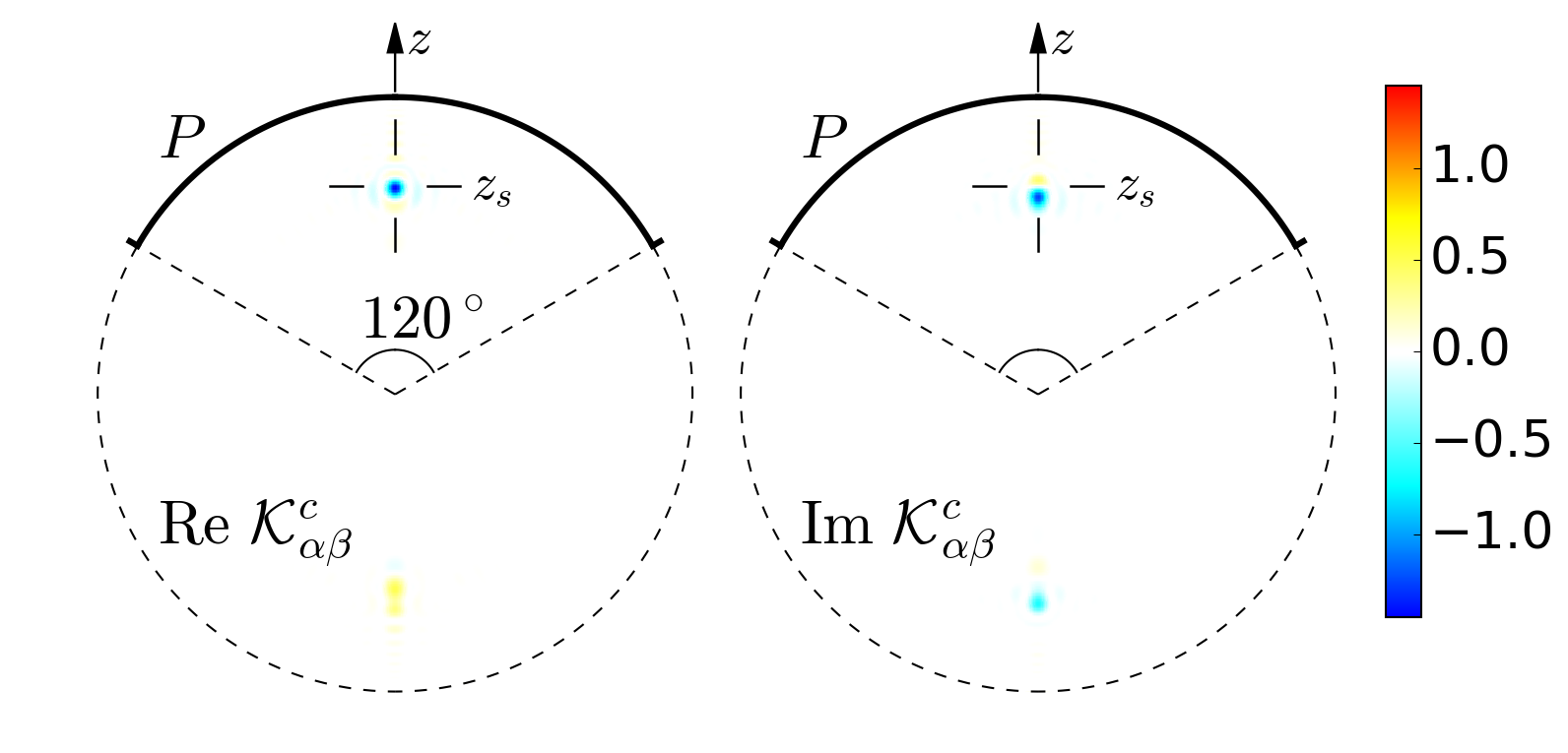}
\caption{Meridional slice through the sound-speed kernel $\mathcal{K}_{\alpha\beta}^c(\bx,\sc)$ computed in Model S at a single frequency of 3~mHz, in units of $10^{-34}$\,kg\,m$^{-3}$\,s$^{-3}$.  Both the real  (\textit{left panel}) and  imaginary  (\textit{right panel}) parts of the kernel are shown. The scatterer at $\zs =0.7\ \mathrm{R_\odot}$ is indicated by the crosses. 
The observation pupil $P$ is a polar cap of full angular size $120^\circ$.
Notice the ghost values at the antipode result from the reflection of waves at the surface due to the rapid drop of the density.}\label{fig.2}
\end{center}
\end{figure}

In order to illustrate the theory, we compute holograms in the presence of sound-speed perturbations at different depths and calculate the corresponding signal-to-noise ratios. 

\subsection{Reference Green's function}
The main input quantity required to compute holograms is the reference Green's function  (Eq.~\ref{Greens}). Here it is computed in the frequency domain using the spherically-symmetric standard solar Model S \citep{CHR96}. The wave attenuation model is taken from \citet{GIZ17}. Below $5.3$~mHz, we have 
$\gamma = \gamma_0 \left|{\omega}/{\omega_0}\right|^{5.77}$,  where $\gamma_0/2\pi=4.29\ \mu\mathrm{Hz}$ and $\omega_0/2\pi=3\ \mathrm{mHz}$. Above $5.3$~mHz,
$\gamma/2\pi = 125 \ \mu\mathrm{Hz}$ is kept constant. The radiation boundary condition defined by Eq.~\eqref{eq:b.c.} is applied at the computational boundary located 500~km above the solar surface with the local wavenumber $k_n$ (where  $H = 105$~km). The wave equation is solved using  the finite-element solver Montjoie \citep{Durufle2006,GIZ17}.

The reference Green's function only depends on the angular distance $\Theta$ between the two points at radii $r$ and $r'$. To speed up the computations, we place one of the points  on the polar axis and compute the axisymmetric component of the Green's function $G_l^{m=0}(r,r', \omega)$ at each spherical harmonic degree $l$, to obtain:
\begin{equation}
G_0(\br,\br', \omega)
\simeq \sum_{l=0}^{l_{\rm max}} 
G_l^{m=0}(r,r', \omega) P_l(\cos\Theta) , \label{eq:G}
\end{equation}
where we use an approximate equality because  the sum is truncated at $l_{\rm max} = 300$. 

\subsection{Sound-speed kernels}

The sound-speed kernel is computed using Eq.~(\ref{eq.Kc}) and the definition of $\mathcal{K}^k_{\alpha\beta}$. One needs to evaluate two surface integrals, which can be computed analytically via a decomposition of all quantities into spherical harmonic coefficients \citep{FOU18}.

\begin{figure}[t]
\begin{center}
\includegraphics[width=\linewidth]{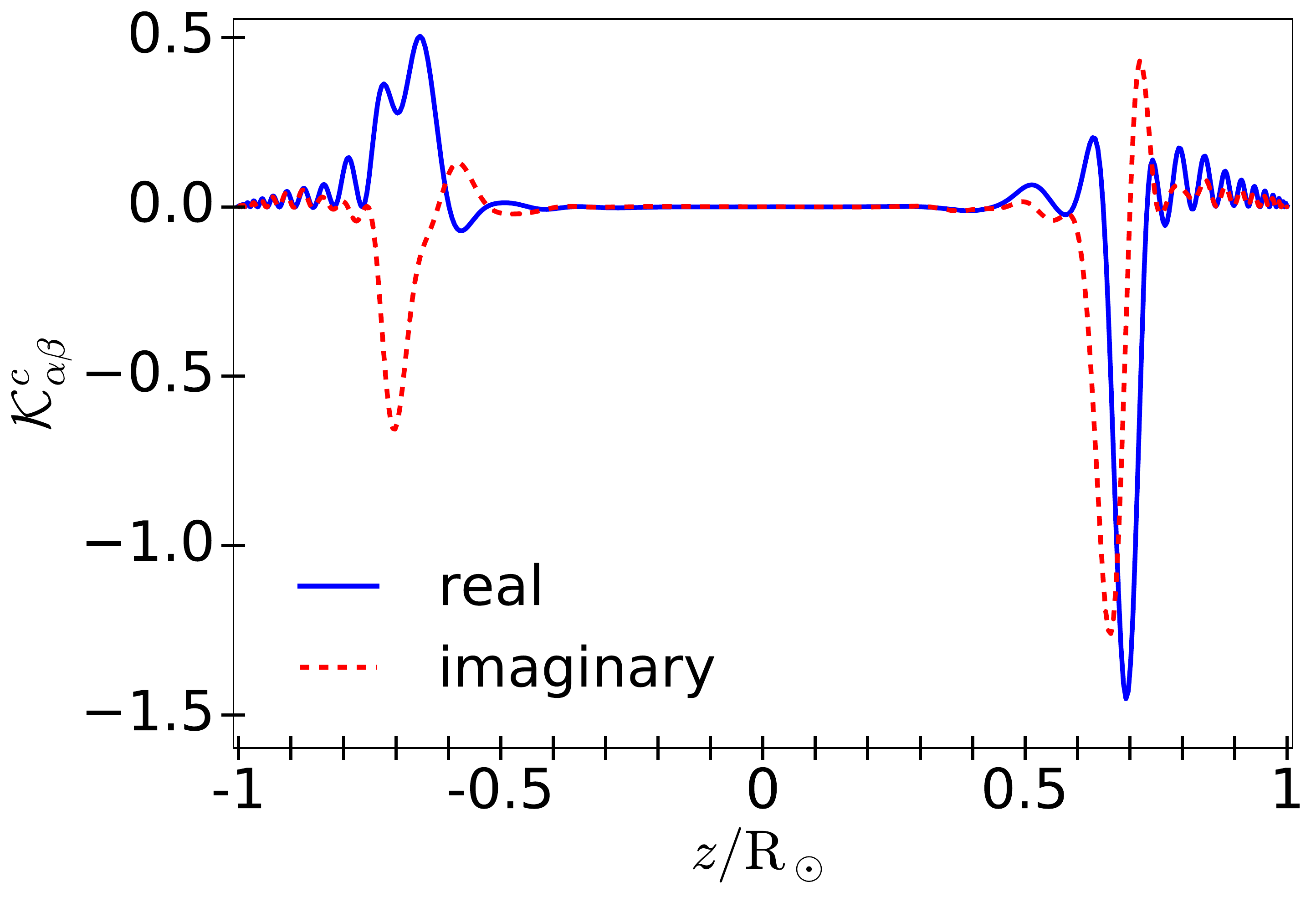}
\caption{Cut along the $z$~axis through the sound-speed kernel from Fig.~\ref{fig.2}. The scatterer is at $\zs =0.7$~$\mathrm{R_\odot}$.}
\label{fig.3}
\end{center}
\end{figure}

Figure~\ref{fig.2} shows a sound-speed kernel $\mathcal{K}_{\alpha\beta}^c$ at a single frequency of 3~mHz. The pupil $P$ is a polar cap of angular size $120^\circ$ and the wave propagators $H_\alpha$ and $H_\beta$ are given in Table~\ref{tab:pupils}.  As expected the kernel peaks around the scatterer position at $z=0.7\mathrm{R_\odot}$. A cut along the polar axis is shown in Fig.~\ref{fig.3}; the kernel width is about half the local wavelength. In addition, we find ghost values at the antipode.

\subsection{Signal} 
\label{sect:signal}

\begin{figure*}
\begin{center}
\includegraphics[width=\linewidth]{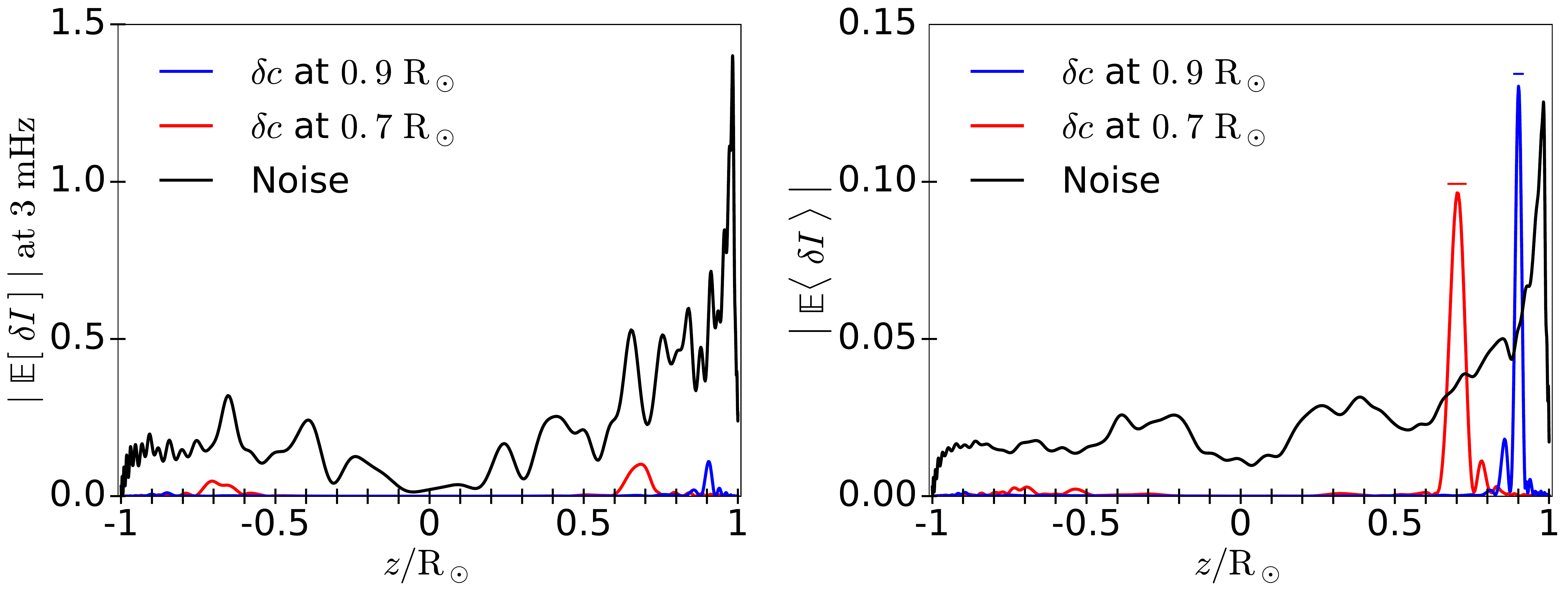}
\caption{Left panel: Hologram intensity $\left|  \EE [ \delta I_{\alpha\beta}(\bx)  ]  \right|$ at a single frequency of 3~mHz, displayed along the $z$-axis (at $\bx= z \mathbf{\hat{z}}$).
The sound-speed perturbation (see Eq.~\ref{eq:delta_c}) is placed at two different positions along the $z$-axis, $\zs = 0.7 \, {\rm R}_\odot$ (\textit{red}) and $0.9 \,{\rm R}_\odot$ (\textit{blue}). 
The standard deviation of the noise  $\sqrt{ {\rm Var}  \langle I_{\alpha\beta, 0}  (\bx)  \rangle }$ is given by the \textit{black} curve.    
Note that the jagged aspect of the curves is not due to numerical inaccuracies. 
Right panel: hologram intensity and noise level after averaging over 101 frequencies uniformly distributed in the interval from $2.75$ to $3.25$~mHz. The frequency resolution is $5$~$\mu$Hz, implying an observation duration of $T =  55.5$~h. A horizontal line segment is plotted at each depth to mark half of the local wavelength.
}\label{fig.4}
\end{center}
\end{figure*}

At position $\sc = \zs  \mathbf{\hat{z}}$ along the polar axis, we consider a localized increase in sound speed of 10\%  over a volume $V_{\rm s}$, such that the signal (Eq.~\ref{eq:signalC}) may be written as
\begin{equation}
\EE [\delta I_{\alpha\beta}(\bx)] \simeq 0.1 V_{\rm s} \, c_0(\sc) \,\mathcal{K}^c_{\alpha\beta}(\bx,\sc). \label{eq:delta_c}
\end{equation}
The volume $V_{\rm s}$ is that of a ball of diameter $\lambda(\rs)/2=\pi/[{\rm Re}\ k(\rs,\omega_0)]$ with $\omega_0/2\pi = 3\ \mathrm{mHz}$. This is an approximate but much faster way to compute the effect of a perturbation of volume comparable to the highest possible holographic resolution. 
It has been checked that the answer does not differ significantly from the one obtained by integrating numerically the kernel over the ball of volume $V_{\rm s}$.
For reference, note that $\lambda/2=38$~Mm at $r = 0.7\ \mathrm{R_\odot}$ and $\lambda/2=20$~Mm at $r=0.9\ \mathrm{R_\odot}$. 

Figure~\ref{fig.4} shows the signal  $\left|  \EE \ [ \delta I_{\alpha\beta}(\bx)  ]  \right|$ for sound-speed perturbations located at two different depths $\zs = 0.7\ \mathrm{R_\odot}$ (red curve) and $0.9\ \mathrm{R_\odot}$ (blue curve). The pupil $P$ and the wave propagators are the same as those  of Fig.~\ref{fig.2}. The left panel of Fig.~\ref{fig.4} shows the results at a single frequency of 3~mHz.  With only one frequency, the signal peaks close to the scattering location but the spatial resolution is relatively poor, with a ghost on the far side. To demonstrate the benefits of averaging, the right panel shows the signal after averaging over 101 frequencies uniformly distributed in the interval $2.75$\,--\,$3.25$~mHz. The frequency resolution $5$~$\mu$Hz corresponds to an observation duration $T=55.5$~h.  Averaging over frequencies improves the spatial resolution which approaches $\lambda/2$ and  the ghost is suppressed. A horizontal line segment is plotted on the right panel at each depth to mark half the local wavelength. 

\begin{figure}[t]
\begin{center}
\includegraphics[width=\linewidth]{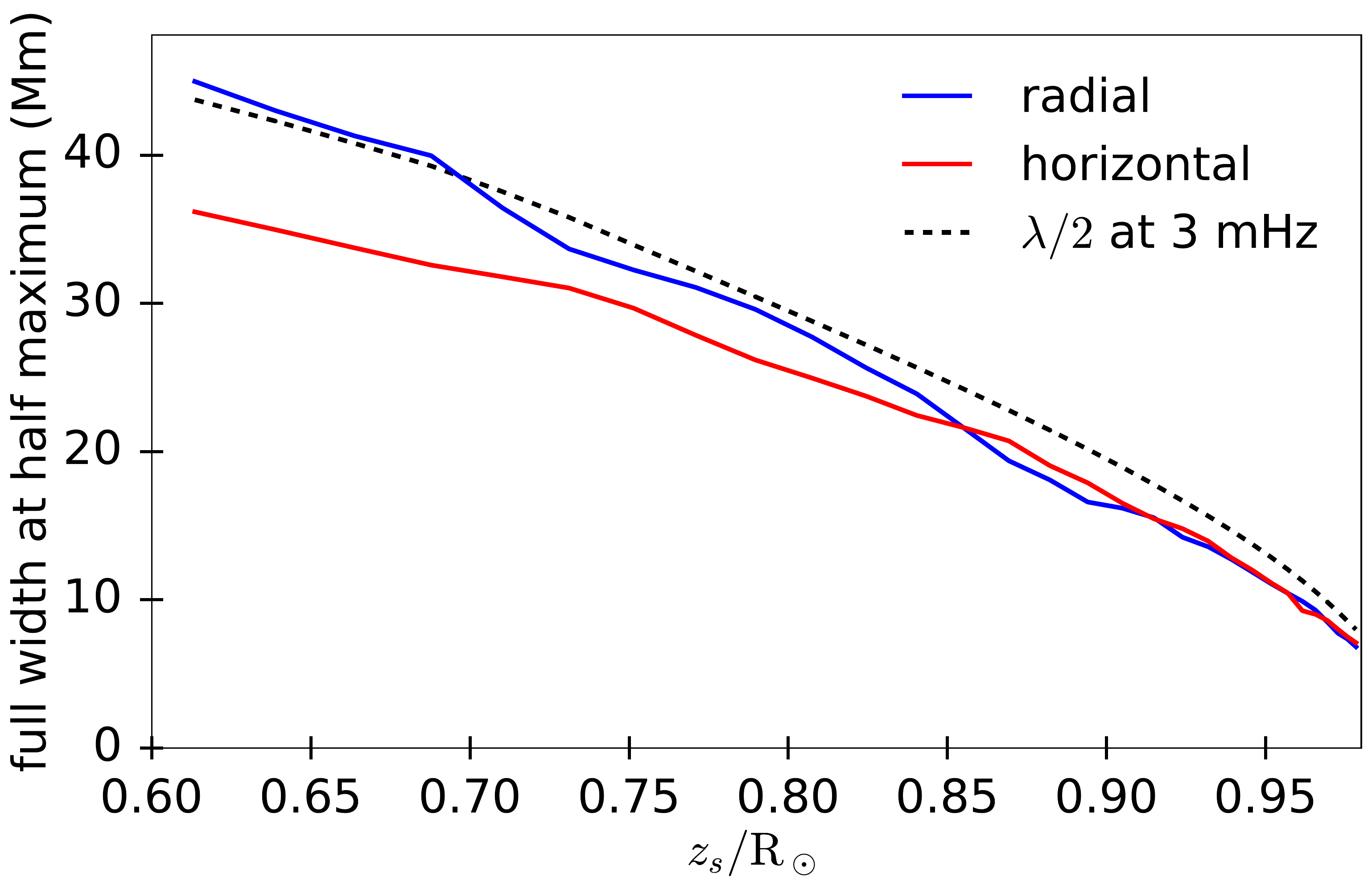}
\caption{Radial and horizontal widths of the frequency-averaged sound-speed kernel $|\langle \mathcal{K}^c \rangle|$ as functions of scattering position. These are close to half the local wavelength at 3~mHz (dotted line). }\label{fig.5}
\end{center}
\end{figure}

As seen in Fig.~\ref{fig.5}
the spatial extent of the frequency-averaged kernels is approximately $\lambda/2$ in both the radial and horizontal directions, for all scattering points in the range $0.6<z_s/\mathrm{R_\odot}<0.98$.
Thus helioseismic holography is a diffraction-limited imaging technique as suggested by \citet{LIN97}.

\subsection{Noise}
The noise is obtained from Eq.~(\ref{eq:noise}), which requires the computation of $\EE  [I_{\alpha,0}]$ and $\EE  [I_{\beta,0}]$ using Eq.~\eqref{eq:I0}. The reference cross-covariance $C_0$ is precomputed. The double surface integral is evaluated in a similar way as for the kernel computations. 

For a frequency of 3 mHz the left panel of Fig.~\ref{fig.4} (black curve) shows the noise level, together with the signal described in the previous section. The jagged aspect of the noise variations with position is not due to numerical inaccuracies but to the details of the Green's function. As shown in the right panel of Fig.~\ref{fig.4}, the noise level goes down by a factor of about ten after averaging over 101 frequencies, and varies more smoothly with depth.     

{
\citet{BRA08} studied the noise level in observed travel times measured from LB holography. These measurements, however, include contributions from supergranulation and so are not directly comparable to what is shown in Fig.~\ref{fig.4}.
}

\subsection{Signal-to-noise ratio}

\begin{figure}
\begin{center}
\includegraphics[width=\linewidth]{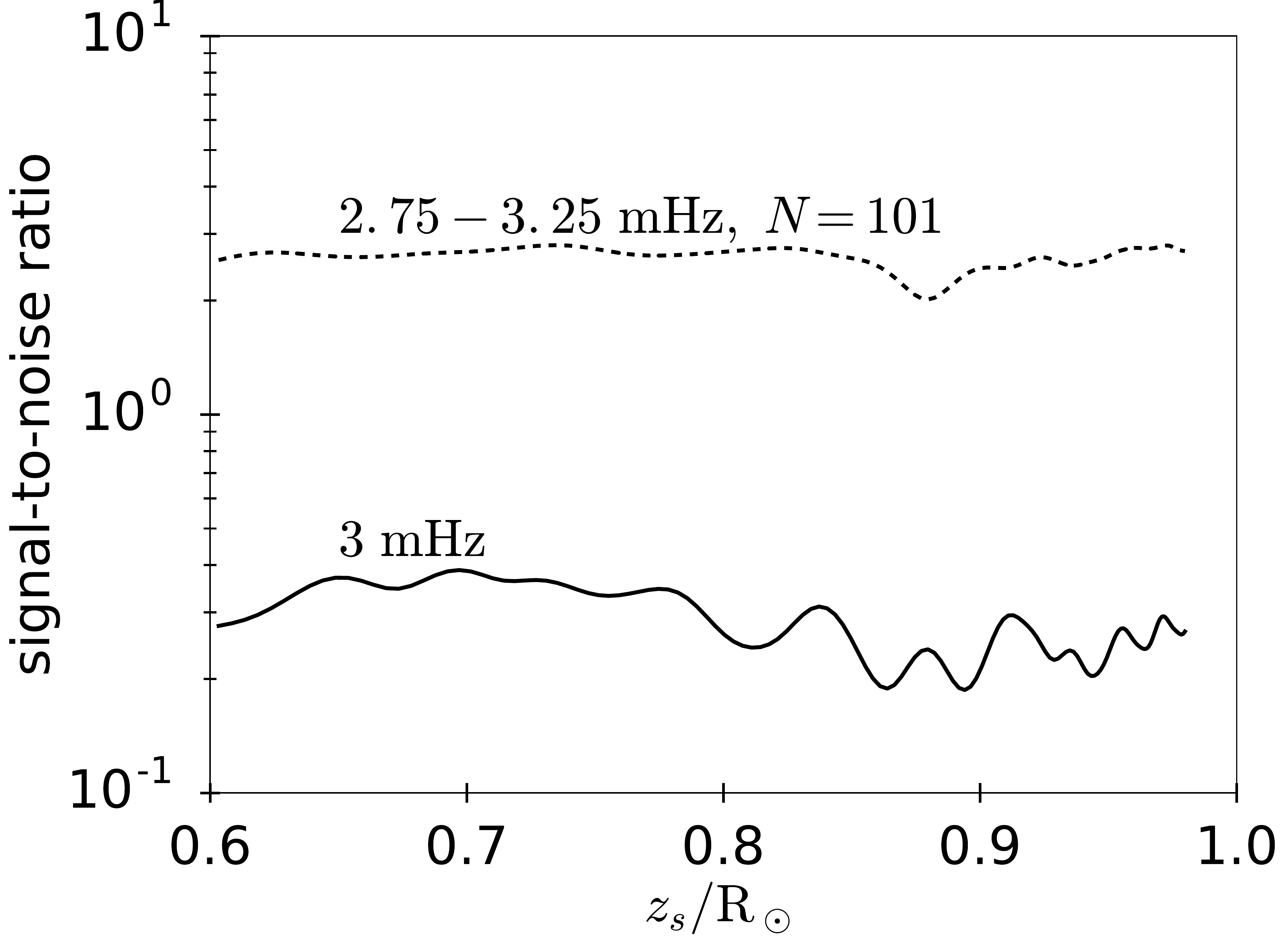}
\caption{Signal-to-noise ratio in hologram intensity for a 10\% sound-speed perturbation over a volume $V_{\rm s}(\zs)$  placed at $z_{\rm s}$ along the polar axis (Eq.~\ref{eq:delta_c}). The results are shown at a single frequency of 3~mHz (\textit{solid}) and after averaging over 101 frequencies in the interval from $2.75$ to $3.25$~mHz (\textit{dashed}).  
}\label{fig.6}
\end{center}
\end{figure}

Figure~\ref{fig.6} shows the signal-to-noise ratio as a function of scatterer location. We recall that the sound-speed perturbation is specified by Eq.~\eqref{eq:delta_c} and is the same as in Sect.~\ref{sect:signal}. The results are shown at a single frequency of 3~mHz and after averaging over 101 frequencies  in the interval $2.75$\,--\,$3.25$~mHz.  After averaging, the signal-to-noise ratio is above 2 and is roughly independent of depth in the range $0.6 < \zs/{\rm R}_\odot<0.98$ for a pupil of angular size $120^\circ$. Note that the ghost at $-\zs$ is much below the noise level.

\begin{figure}
\begin{center}
\includegraphics[width=\linewidth]{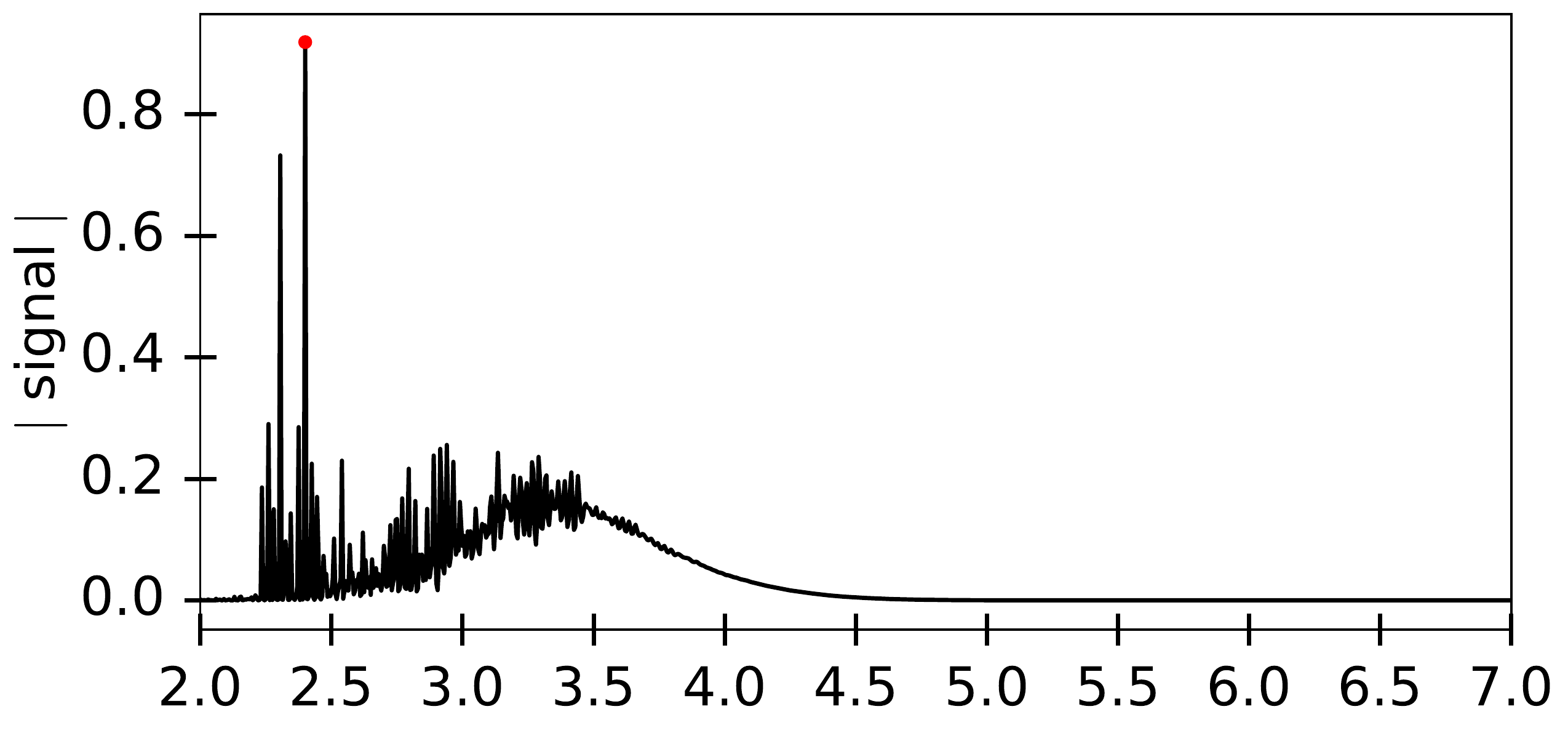}\\
\includegraphics[width=\linewidth]{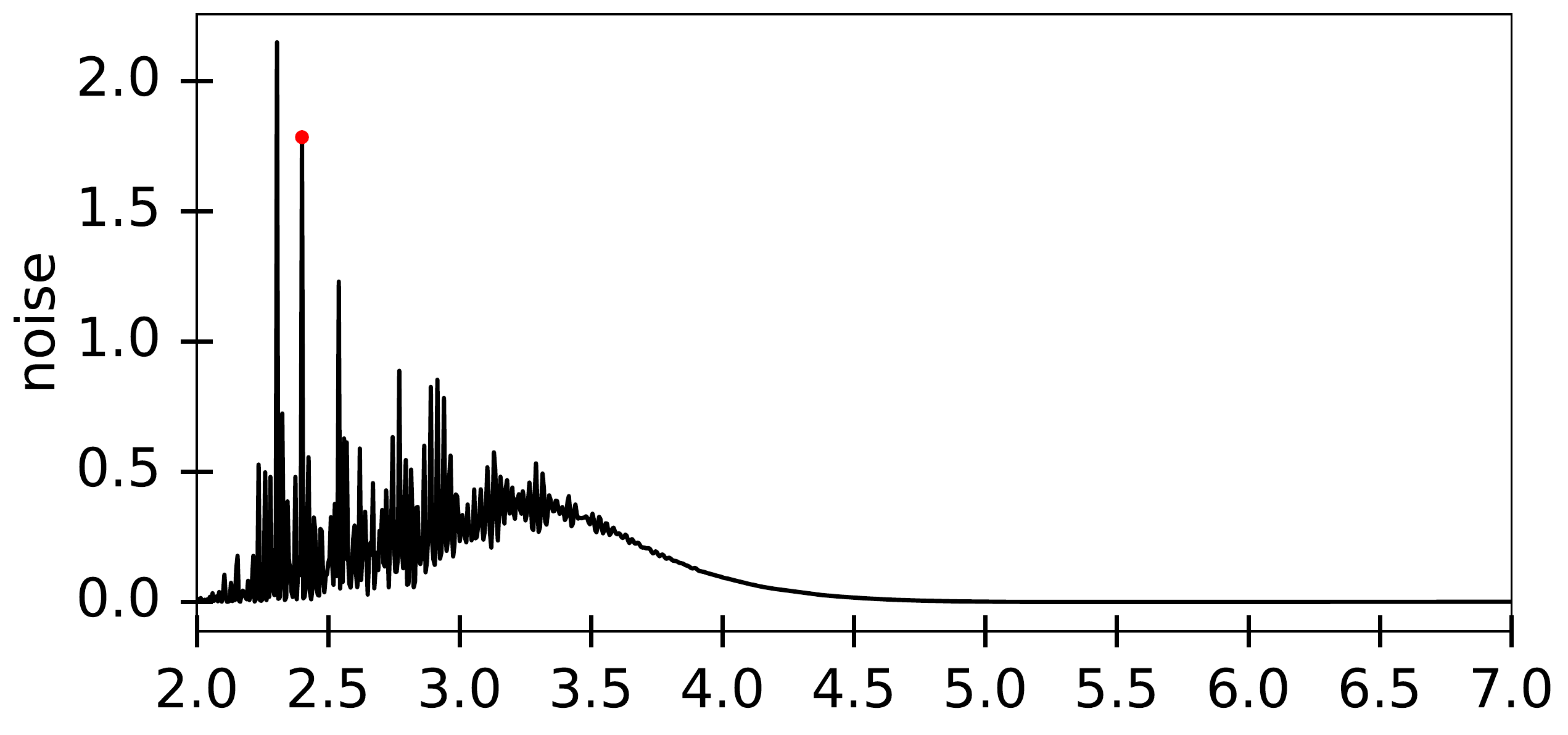}\\
\includegraphics[width=\linewidth]{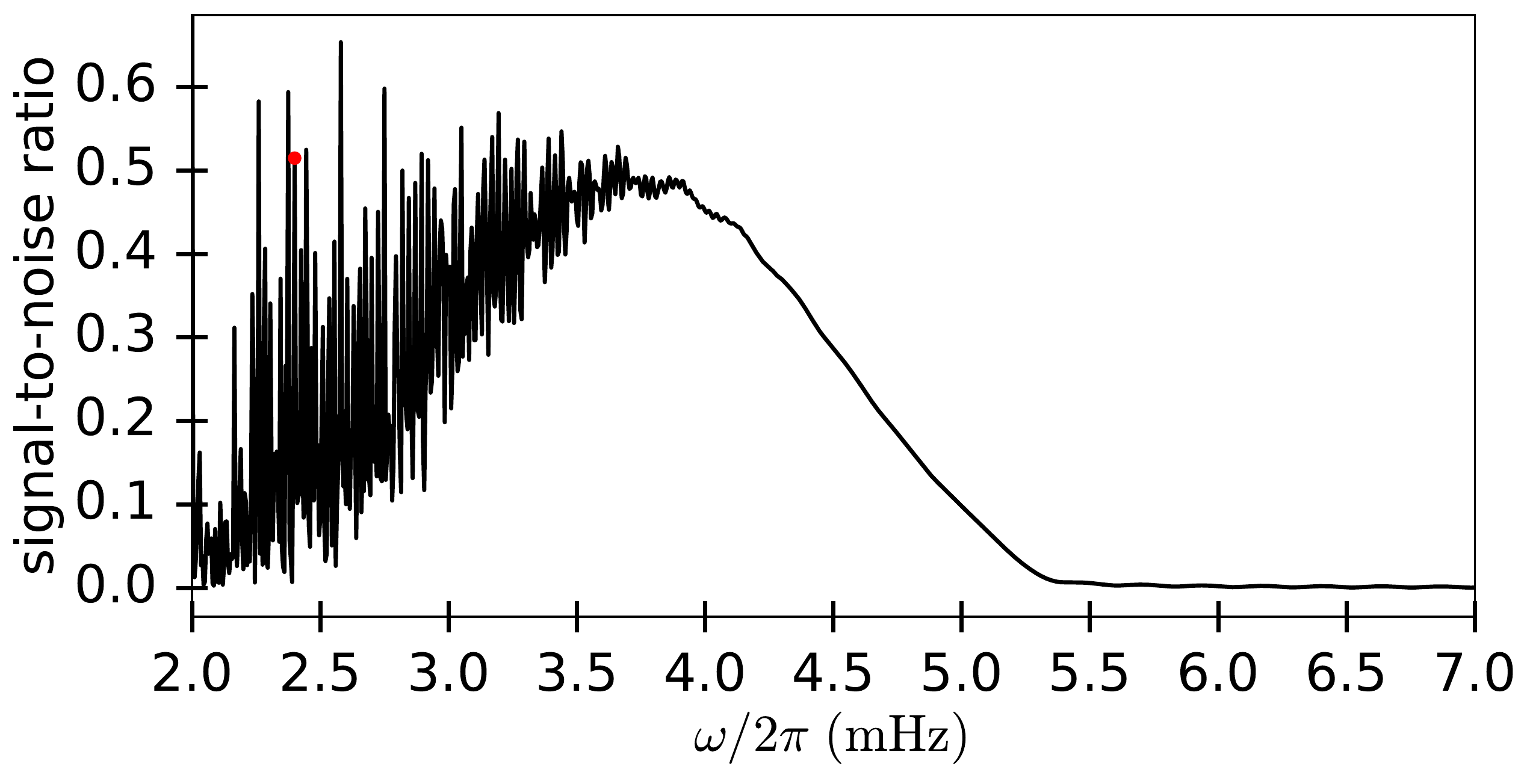}
\caption{Signal, noise, and signal-to-noise ratio as function of frequency for a sound-speed scatterer located at $\zs = 0.7\ \mathrm{R}_\odot$. 
Here we show the result for a frequency range of 2 to 7~mHz. The rapid changes are not due to numerical inaccuracies. The red dots mark the spikes in the signal and the noise at $2.4000\ \mathrm{mHz}$.}
\label{fig.7}
\end{center}
\end{figure}

\begin{figure}
\begin{center}
\includegraphics[width=\linewidth]{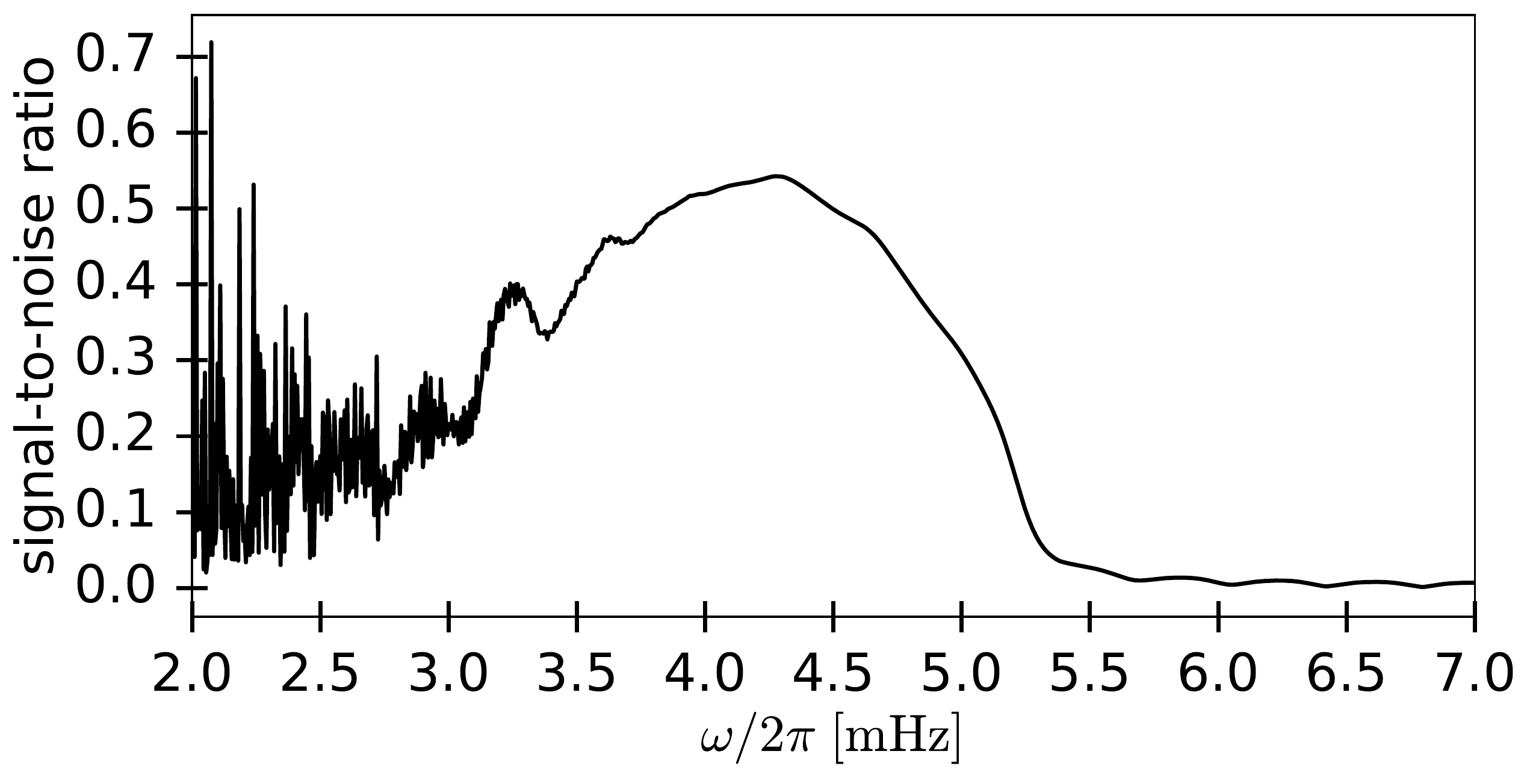}
\caption{{Signal-to-noise ratio as in Fig.~\ref{fig.7}, but for a sound-speed scatterer located closer to the surface at $\zs = 0.9\ \mathrm{R}_\odot$.}}
\label{fig.8}
\end{center}
\end{figure}

We find that both signal and noise vary rapidly with frequency for deep located sound-speed scatterers. Figure~\ref{fig.7}  shows an example of a sound-speed scatterer located at $\zs = 0.7 \ \mathrm{R}_\odot$. Strong frequency variations in signal and noise are evident for frequencies below 3.5~mHz. This can be understood as follows. Low-frequency modes have narrowly-peaked power spectra due to their long lifetimes. At these low frequencies, the amplitude of the kernel function and the noise will change rapidly when the frequency coincides with a particular mode frequency. Additionally, the kernel function may not peak exactly at the sound-speed scatterer position when only a few modes contribute to the kernel function. {Figure~\ref{fig.8}  shows the signal-to-noise ratio as a function of frequency for a sound-speed scatterer located at $\zs = 0.9\ \mathrm{R}_\odot$. For this target depth closer to the surface, the rapid variations disappear above 3~mHz, due to the larger contribution of high-degree modes which are not resolved in frequency space because of their short lifetimes.
}

\begin{figure}
\begin{center}
\includegraphics[width=\linewidth]{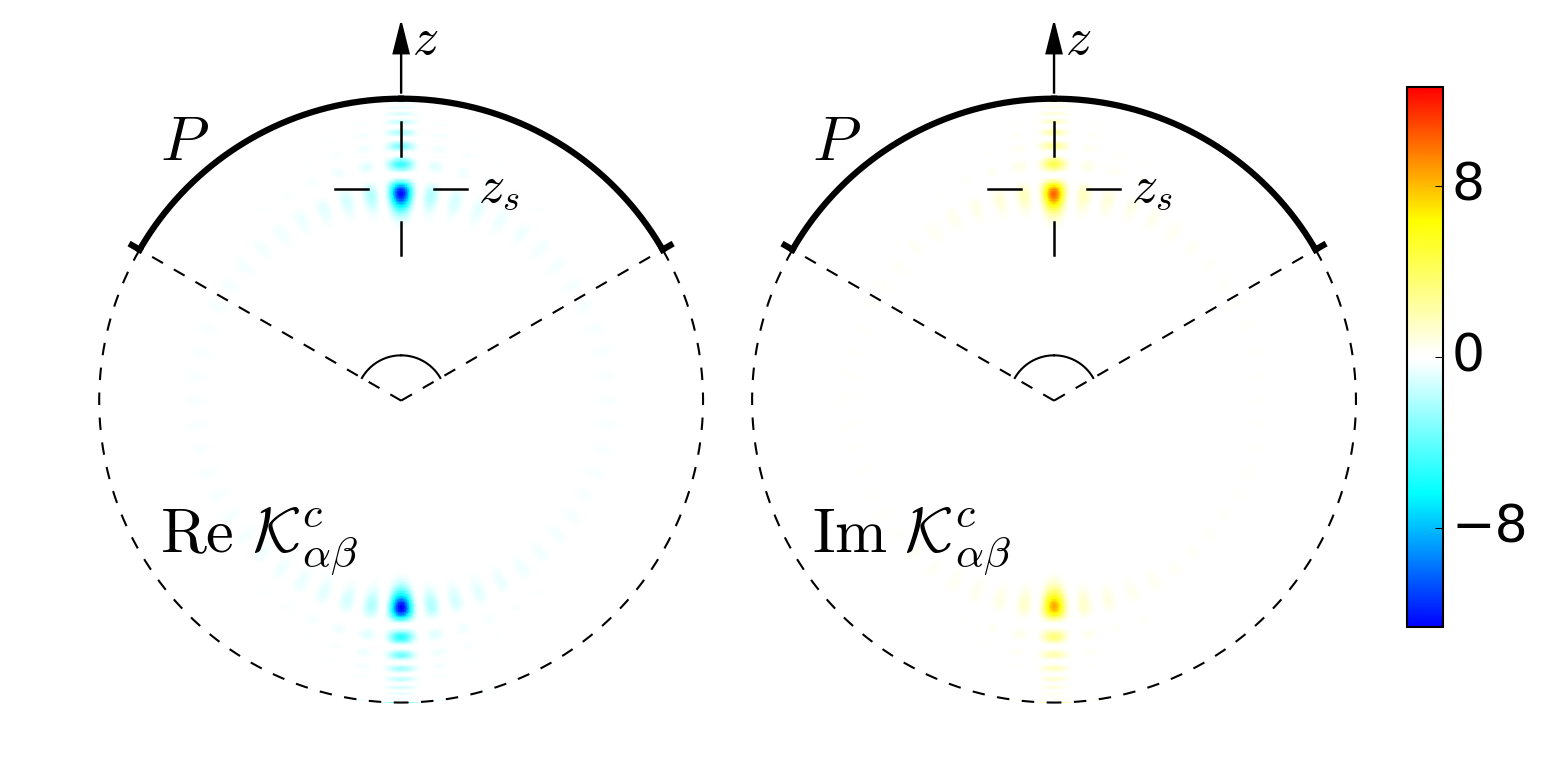}
\caption{Meridional slice of the sound-speed kernel $\mathcal{K}_{\alpha\beta}^c$ with $z_{\rm s} = 0.7\ \mathrm{R}_\odot$ computed at the low frequency of $2.4000$~mHz, which corresponds to the spike with a red dot in Fig.~\ref{fig.7}. This kernel displays oscillations and is not peaked  as much as the 3-mHz kernel from Fig.~\ref{fig.2}. }
\label{fig.9}
\end{center}
\end{figure}

The kernel function  at frequency $2.4000$~mHz {for $\zs=0.7R_\odot$} is shown in Fig.~\ref{fig.9}; this particular frequency corresponds to the peak marked in Fig.~\ref{fig.7} with a red dot. We see that this kernel is much less localized around the scattering point than the kernel at 3 mHz (Fig.~\ref{fig.2}).

\section{Conclusion}
We derived a framework for computing the expected signal and the noise level in PB helioseismic holography. The same framework could be used to interpret LB holograms and phase-sensitive holograms.

PB holography requires knowledge of the wave field, $\psi=\rho^{1/2} c^2 \nabla\cdot\bxi$, and its normal derivative, $\partial_n \psi$, on the solar surface.  
With this definition of $\psi$, the Green's function used in the definition of PB holograms solves a well-defined  Helmholtz-like equation, which we solve numerically  \citep{GIZ17}. The need for finite-frequency Green's functions was demonstrated in LB holography by \citet{Perez2010}.
In the numerical examples shown in the previous section, we assumed that we have full knowledge of $\partial_n \psi$ on the surface. In practice, we do not observe directly the normal derivative of the wave field; it must be approximated. According to complementary simulations (not shown here), this can be achieved by using the approximate outgoing radiation condition $\partial_n \psi = \ii k_n \psi$  derived by  \citet{BAR17}.

We found that, for a sufficiently large pupil, scatterers can be imaged at a resolution that is very close to half the local wavelength, $\lambda/2$. This confirms the claim by \citet{LIN97,LIN00a}  that helioseismic holography is  diffraction-limited. In that sense, helioseismic holography is superior to deep-focusing time-distance helioseismology, which gives lower spatial resolution \citep{Munk2001,Pour2018}.

For large pupils, we found that the signal-to-noise ratio in holograms does not vary much with depth in the convection zone, when a perturbation in sound-speed fills a volume corresponding to the holographic resolution.

Averaging over frequencies improves the signal-to-noise ratio. For a scatterer at the bottom of the convection zone, the signal and the noise vary smoothly with frequency above $4$~mHz (see Fig.~\ref{fig.7}). At lower frequencies, however, the signal varies rapidly with frequency (due to contributions from individual long-lived p~modes)
and it is not obvious how the signal should be averaged.  A specific analysis of low-frequency holograms is required, especially for deep scatterers.

We found that the signal-to-noise  ratio in PB holography is maximum around $3.7$~mHz for $\zs=0.7$\ R$_\odot$
(resp. at $4.3$~mHz for $\zs=0.9$\ R$_\odot$). There is no indication in our calculations that there is a benefit in using the frequencies above the acoustic cutoff \citep[unlike predictions by][for phase sensitive holography]{Ruz2003}. {The signal-to-noise ratio drops to very small values above 5~mHz. One may ask if this drop is somehow compensated by the increase in spatial resolution at high frequencies. The answer is negative. Our calculations indicate that noise has a horizontal correlation length that is about half the local wavelength. Far too few independent measurements are available at high frequencies to recover a decent signal-to-noise ratio by horizontal spatial averaging.}

Our synthetic data do not contain a convective background. The effect of this background on signal-to-noise ratios in holography should be studied. 
Future work should also investigate the performance of PB holography for target locations that are away from the axis of the pupil, especially for farside imaging applications. 

\begin{acknowledgements}
The theoretical framework was developed by L.G. and D.F. at Mathematisches Forchungsinstitut Oberwolfach in May 2017. The numerical computations were performed by D.Y. using the Montjoie solver. D.Y. is a member of the International Max Planck Research School for Solar System Science at the University of G\"ottingen. We thank M.~Durufl\'e and J.~Chabassier from Inria Magique-3D for the helioseismology-related developments of Montjoie. We also thank Chris~Hanson from NYUAD for the fine tuning of the model power spectrum of solar oscillations. L.G. acknowledges support from NYUAD Institute grant G1502. The computing resources were provided in part by the German Data Center for SDO, a project funded by the German Aerospace Center (DLR).
\end{acknowledgements}

\bibliographystyle{aa}
\bibliography{biblio}

\end{document}